\documentclass[aps,prl,reprint,superscriptaddress,
amsfonts,amssymb,amsmath,a4paper,floats,floatfix]{revtex4-1}
\usepackage{graphicx}
\usepackage{siunitx}
\usepackage[dvipsnames]{xcolor}
\usepackage{xspace}
\usepackage[utf8]{inputenc}
\usepackage{amsmath}

\begin{document}


\newcommand{\TDEG}{two-dimensional electron gas}
\newcommand{\mos}{MoS\ensuremath{_{\mathrm{2}}}\xspace}
\newcommand{\mose}{MoSe\ensuremath{_{\mathrm{2}}}\xspace}
\newcommand{\wse}{WSe\ensuremath{_{\mathrm{2}}}\xspace}
\newcommand{\ws}{WS\ensuremath{_{\mathrm{2}}}\xspace}
\newcommand{\unit}[1]{\ensuremath{\,\mathrm{#1}}}
\newcommand{\sub}[1]{\ensuremath{_{\mathrm{#1}}}}
\newcommand{\super}[1]{\ensuremath{^{\mathrm{#1}}}\xspace}
\newcommand{\psqcm}{\ensuremath{\,\mathrm{cm}^{-2}}\xspace}
\newcommand{\VTG}{\ensuremath{V_{\mathrm{TG}}}\xspace}
\newcommand{\VBG}{\ensuremath{V_{\mathrm{BG}}}\xspace}
\newcommand{\me}{\ensuremath{\,m_\mathrm{e}}\xspace}
\newcommand{\effm}{\ensuremath{\,m^\mathrm{\ast}}\xspace}
\newcommand{\tmd}{TMD\xspace}
\newcommand{\tmds}{TMDs\xspace}
\newcommand{\mev}{\ensuremath{\,\mathrm{meV}}\xspace}
\newcommand{\FigRef}[1]{Fig.$\,$\ref{#1}}
\newcommand{\FigRefi}[2]{\FigRef{#1}$\,$(#2)}
\newcommand{\emob}{\ensuremath{\,\mathrm{cm}^2/\mathrm{V}\mathrm{s}}\xspace}
\newcommand{\kb}{\ensuremath{\,k_\mathrm{B}}\xspace}
\newcommand{\hwc}{\ensuremath{\,\hbar\omega_\mathrm{c}}\xspace}
\newcommand{\Um}{\ensuremath{\,\mu\mathrm{m}}\xspace}
\newcommand{\DiB}{{\ensuremath D_\mathrm{B}}}
\newcommand{\Dibl}{{\ensuremath D_\mathrm{BL}}}
\newcommand{\CT}{{\ensuremath C_\mathrm{T}}}
\newcommand{\CB}{{\ensuremath C_\mathrm{B}}}
\newcommand{\Cbl}{{\ensuremath C_\mathrm{BL}}}
\newcommand{\nT}{{\ensuremath n_\mathrm{t}}}
\newcommand{\nB}{{\ensuremath n_\mathrm{b}}}
\newcommand{\dT}{{\ensuremath d_\mathrm{T}}}
\newcommand{\dbl}{{\ensuremath d_\mathrm{BL}}}
\newcommand{\ebl}{{\ensuremath \epsilon_\mathrm{BL}}}
\newcommand{\VT}{{\ensuremath V_\mathrm{TG}}}
\newcommand{\VB}{{\ensuremath V_\mathrm{BG}}}


\title{Absence of inter-layer tunnel coupling of $K$-valley electrons in bilayer \mos}

\author{Riccardo Pisoni}
	\affiliation{Solid State Physics Laboratory, ETH Z\"urich, 8093 Z\"urich, Switzerland}
	
\author{Tim\ Davatz}
	\affiliation{Solid State Physics Laboratory, ETH Z\"urich, 8093 Z\"urich, Switzerland}

\author{Kenji Watanabe}
	\affiliation{National Institute for Material Science, 1-1 Namiki, Tsukuba 305-0044, Japan}

\author{Takashi Taniguchi}
	\affiliation{National Institute for Material Science, 1-1 Namiki, Tsukuba 305-0044, Japan}
	
\author{Thomas\ Ihn}
	\affiliation{Solid State Physics Laboratory, ETH Z\"urich, 8093 Z\"urich, Switzerland}
	
\author{Klaus\ Ensslin}
	\affiliation{Solid State Physics Laboratory, ETH Z\"urich, 8093 Z\"urich, Switzerland}

\date{\today}


\begin{abstract}
In Bernal stacked bilayer graphene  interlayer coupling significantly affects the electronic bandstructure compared to monolayer graphene. 
Here we present magnetotransport experiments on high-quality $n$-doped bilayer \mos. By measuring the evolution of the Landau levels as a function of electron density and applied magnetic field we are able to investigate the occupation of conduction band states, the interlayer coupling in pristine bilayer \mos, and how these effects are governed by electron-electron interactions.
We find that the two layers of the bilayer \mos behave as two independent electronic systems where a two-fold Landau level's degeneracy is observed for each \mos layer.
At the onset of the population of the bottom \mos layer we observe a large negative compressibility caused by the exchange interaction.
These observations, enabled by the high electronic quality of our samples, demonstrate weak interlayer tunnel coupling but strong interlayer electrostatic coupling in pristine bilayer \mos. The conclusions from the experiments may be relevant also to other transition metal dichalcogenide materials.
\end{abstract}

\maketitle


Of the multitude of two-dimensional (2D) host materials, transition metal dichalcogenides (\tmds) are promising candidates for exploring quantum correlated electronic phases and electron-electron interaction effects due to their intrinsic 2D nature, large spin-orbit interaction and large effective mass carriers. 
Molybdenum disulfide (\mos) is one of the most widely studied \tmds and still most of its fundamental quantum electronic properties have thus far been elusive.
Contrary to monolayer \mos, in pristine bilayer \mos inversion symmetry is restored~\cite{xiao_coupled_2012,mak_control_2012,cao_valley-selective_2012}. 
As a result, the orbital magnetic moment and the valley-contrasting optical dichroism vanish~\cite{yao_valley-dependent_2008,xiao_coupled_2012}.
A potential difference between the two layers breaks the inversion symmetry  ~\cite{wu_electrical_2013,lee_electrical_2016}.
The  influence of a perpendicular electric field on bilayer \mos has been extensively probed by optical excitation~\cite{wu_electrical_2013,lee_electrical_2016}.
Very little is known about the electronic transport properties of bilayer \mos when electric and magnetic fields are both applied perpendicular to the sample plane~\cite{lin_probing_2018}.
Magnetotransport studies of 2D holes have been recently performed in bilayer \wse revealing the presence of two subbands, each localized in the top and bottom layer, and demonstrating an upper bound of the interlayer tunnel coupling of $19\mev$~\cite{fallahazad_shubnikovchar21haas_2016, movva_density-dependent_2017}.
A thorough study of the interlayer coupling in the conduction band of bilayer transition metal dichalcogenides is still missing~\cite{lin_probing_2018, larentis_large_2018}. 

Here we report a magnetotransport study of electrons in the conduction band of dual-gated bilayer \mos. All studied bilayer samples exhibit Shubnikov-de Haas (SdH) oscillations with a twofold Landau level degeneracy at $T = 1.5\unit{K}$. At lower temperature the valley degeneracy is lifted and spin-valley coupled Landau levels are resolved. The evolution of the Landau level spectrum as a function of density indicates that electrons occupy states of the $K$ and $K'$ valley in each layer.
By tuning the Fermi energy in each layer individually we are able to populate lower and upper spin-orbit split bands in both layers.
The exchange interaction in a single layer yields a pronounced negative compressibility visible in occupation of the states detected via the Landau fan diagram.
In addition, we observe an intricate interplay between spin- and valley-polarized Landau levels originating from the two decoupled \mos layers.
We do not observe any obvious signature in the Landau level spectrum when the electrostatic potential difference between the two layers vanishes and the structural inversion symmetry is expected to be restored.


\begin{figure}[t!]
\includegraphics[width=1\columnwidth]{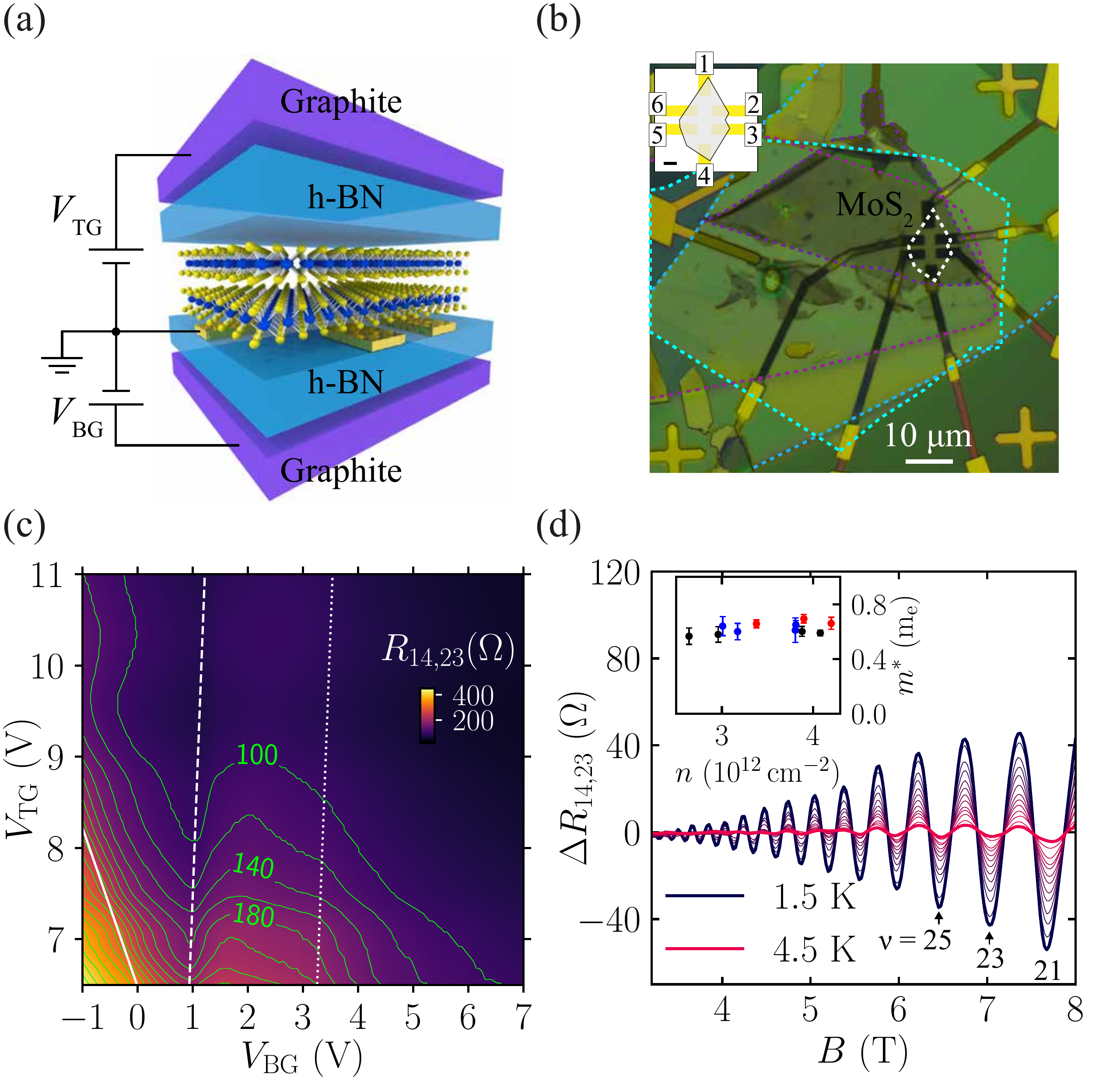}
\caption{\label{fig:fig1}
(a)~Schematic of the device. Bilayer \mos is encapsulated between two hBN layers and Au contacts are pre-patterned on the bottom hBN before the \mos layer is transferred. Graphite flakes serve as bottom and top gates.
(b)~Optical micrograph of sample A. The bilayer \mos flake is highlighted with a white dashed line. Inset: contact geometry, current is injected to contact 1 and extracted from contact 4, voltage is measured between contacts 2 and 3 (scale bar is $2 \Um$).
(c)~Four-terminal resistance $R\sub{14,23}$ as a function of \VTG and \VBG at $T=1.5\unit{K}$. Equi-resistance lines are highlighted in green. The white dashed line denotes the \VBG values at which the bottom \mos layer starts to be populated. Solid and dotted lines represent \VTG and \VBG values at which the upper spin-orbit split bands are occupied in the top and bottom \mos layers, respectively.
(d)~Temperature dependence of the SdH oscillations at $\VBG = -0.4\unit{V}$ and $\VTG = 9\unit{V}$, $n \approx 3.8 \times 10^{12} \psqcm$. An odd filling factor sequence $\nu = 21, 23, 25, ...,$ is observed. Inset: electron effective masses (\effm) extracted for the three different samples as a function of electron density. Black, blue, and red markers represent samples A, B, and C, respectively.
}
\end{figure}

Figure~\ref{fig:fig1}(a) shows the schematic cross-section of the dual-gated bilayer \mos device under study.
The \mos is encapsulated between two hBN dielectrics with graphite layers as top and bottom gates. We fabricate pre-patterned Au bottom contacts below the bilayer \mos. 
Ohmic behavior of these contacts is achieved by applying a sufficiently positive top gate voltage (\VTG).
The heterostructure is assembled using a dry pick-up and transfer method~\cite{pisoni_gate-defined_2017,pisoni_gate-tunable_2018,pisoni_interactions_2018}.
We fabricated and measured three bilayer \mos samples, labeled A, B, and C, which show the same behavior. We will mainly discuss samples A and B here.

The samples use commercial bulk \mos crystal (SPI supplies) mechanically exfoliated on SiO\sub{2}/Si substrates. Using a combination of optical contrast, photoluminescence spectroscopy and atomic force microscopy bilayer \mos flakes are identified.
Figure~\ref{fig:fig1}(b) shows the optical micrograph of sample A. The bilayer \mos flake is outlined with a white dashed line. Top and bottom graphite gates are outlined in purple and top and bottom hBN are outlined in blue and cyan, respectively. In the inset of \FigRefi{fig:fig1}{b} we sketch the contact geometry where contacts 1 and 4 are used for current injection and extraction and contacts 2 and 3 serve as voltage probes.

Figure~\ref{fig:fig1}(c) shows the four-terminal resistance, $R\sub{14,23}$, as a function of \VBG and \VTG at $T=1.5\unit{K}$. Green solid lines denote specific resistance values as a function of \VTG and \VBG. At fixed \VTG we observe a sudden increase in $R\sub{14,23}$ at $\VBG \approx 1\unit{V}$ (white dashed line) that we attribute to the population of the bottom \mos layer. As a result, for the \VBG values on the left side of the white dashed line we probe the electron transport only through the top \mos layer.
At $\VBG \approx 3.4\unit{V}$ (white dotted line) and for \VTG and \VBG values along the white solid line, we observe additional resistance kinks that we attribute to the occupation of the upper spin-orbit split bands in the bottom and top \mos layers, respectively.

We investigate magnetotransport phenomena in bilayer \mos using lock-in techniques at $31.4\unit{Hz}$.
Figure~\ref{fig:fig1}(d) shows $\Delta R\sub{14,23}$, the four-terminal linear resistance with a smooth background subtracted, as a function of magnetic field $B$ at various temperatures ranging from $1.5\unit{K}$ to $4.5\unit{K}$ and a density $n = 3.8 \times 10^{12} \psqcm$. 
For $T=1.5\unit{K}$ SdH oscillations start at $\approx 3\unit{T}$.
At $T=100\unit{mK}$ the onset of SdH oscillations moves to yet lower magnetic fields yielding a lower bound for the quantum mobility of $\approx 5000\emob$ (see \FigRef{fig:fig4}).
The electron density is calculated from the period of the SdH oscillations in $1/B$ considering valley degenerate Landau levels at the $K$ and $K'$ conduction band minima~\cite{pisoni_interactions_2018}.
At $T=1.5\unit{K}$ and $n = 3.8 \times 10^{12} \psqcm$ we observe the sequence of odd filling factors $\nu = 21, 23, 25,... $. The twofold Landau level's valley degeneracy is lifted at lower temperatures (see below).

We determine the electron effective mass \effm from the temperature dependence of the SdH oscillations by fitting $\Delta R\sub{14,23}$ to $\varepsilon/\sinh(\varepsilon)$, where $\varepsilon = 2\pi\super{2}\kb T/\hwc$ and $\hwc = eB/\effm$ is the cyclotron frequency~\cite{ando_electronic_1982,isihara_density_1986,pudalov_probing_2014}.
The inset of \FigRefi{fig:fig1}{c} shows the extracted \effm for the three different samples. We obtain a density-averaged mass of $\effm \approx 0.62\me$ which does not show any obvious dependence neither on $n$ nor on $B$~\cite{zhang_density-dependent_2005,attaccalite_correlation_2002}.

In \FigRefi{fig:fig1}{d} we extract the \effm of the $K$ and $K'$ electrons localized in the top \mos layer, thus effectively calculating the effective mass of a monolayer \mos~\cite{lin_probing_2018}.
The effective masses we extract in our bilayer samples are systematically $10-20 \%$ lower compared to the ones measured in monolayer \mos~\cite{pisoni_interactions_2018}.
In bilayer \mos the top \mos layer is encapsulated between hBN and the bottom \mos layer, which is devoid of electrons. 
We speculate that the higher dielectric constant ($\epsilon\approx 6.4$) reported for monolayer \mos~\cite{cheiwchanchamnangij_quasiparticle_2012, ramasubramaniam_large_2012, molina-sanchez_phonons_2011, laturia_dielectric_2018} compared to hBN ($\epsilon\approx 3.5$) causes a weakening of electron-electron interaction effects thus affecting the \effm value. 

The interaction strength is characterized by the dimensionless Wigner-Seitz radius $r_\mathrm{s}=1/(\sqrt{\pi\,n}a^{\ast}_\mathrm{B})$, where $a^{\ast}_\mathrm{B}=a_\mathrm{B}(\kappa m_\mathrm{e}/m^{\ast})$ is the effective Bohr radius and $\kappa$ the dielectric constant. For the considered electron density range we estimate $r_\mathrm{s}=1.9-10$, placing the system in a regime where interaction effects are important ~\cite{okamoto_spin_1999,shashkin_indication_2001,vakili_spin_2004}. 


\begin{figure}[t!]
\includegraphics[width=1\columnwidth]{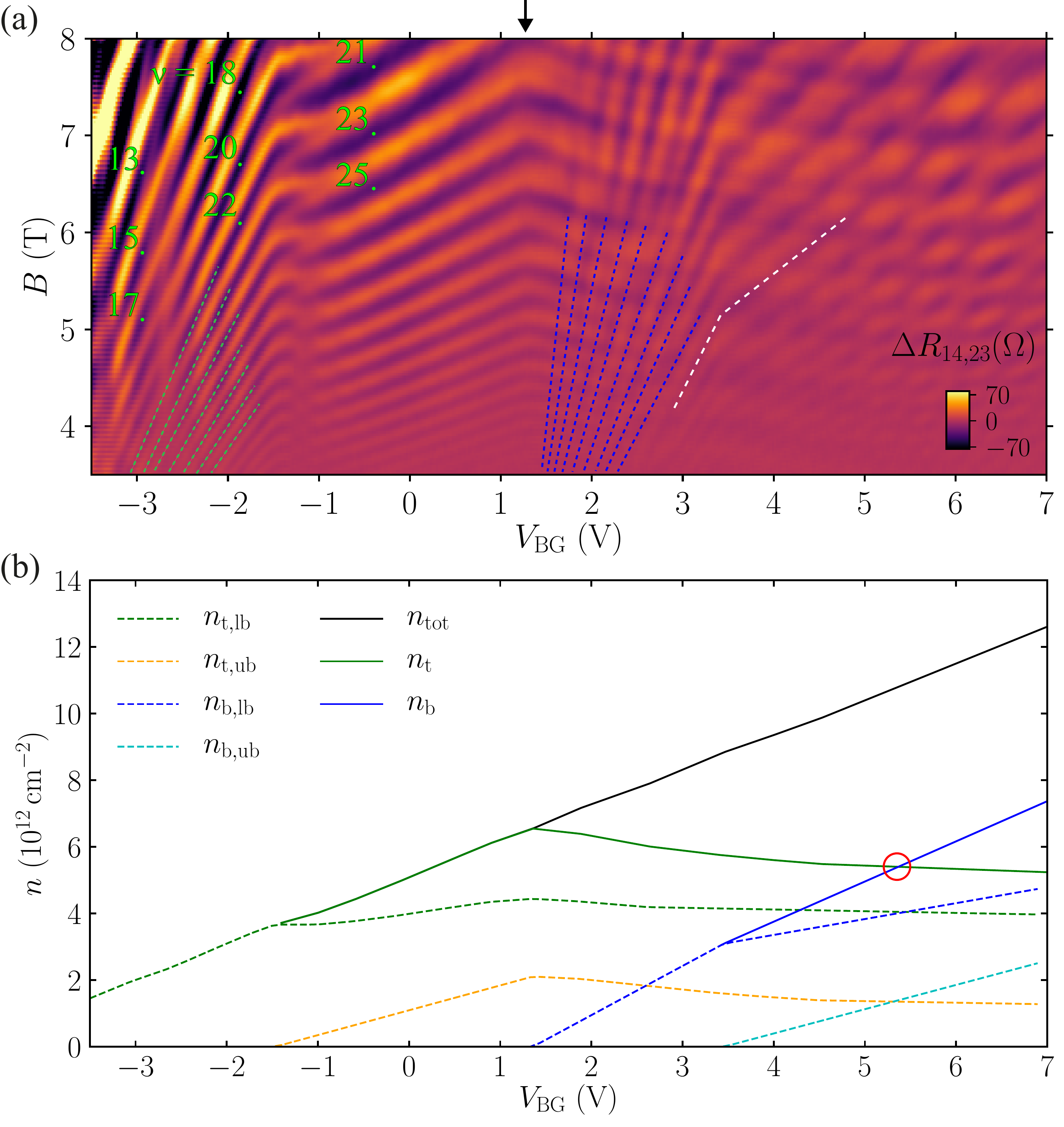}
\caption{\label{fig:fig2}
(a)~Sample A. Four-terminal resistance $\Delta R\sub{14,23}$ as a function of \VBG and magnetic field at $T\approx1.5\unit{K}$ and $\VTG = 9\unit{V}$. For $\VBG < 1.3\unit{V}$ electrons are localized in the top \mos layer. The slope change at $\VBG = -1.5\unit{V}$ indicates the occupation of the upper spin-orbit split bands in the top layer.
At $\VBG = 1.3\unit{V}$ the Landau fan of the bottom \mos layer appears. At $\VBG = 3.4\unit{V}$, the electrons localized in the bottom layer populate the higher energy bands at the $K$ and $K'$ valleys.
(b)~Electron densities in the bilayer \mos bands as a function of \VBG at $T\approx1.5\unit{K}$ and $\VTG = 9\unit{V}$. Green, orange, blue, and cyan dashed lines correspond to electron densities in the lower ($n\sub{t,lb}$, $n\sub{b,lb}$), upper ($n\sub{t,ub}$, $n\sub{b,ub}$) spin-orbit split bands in the top and bottom layer, respectively. Green and blue solid lines represent the total carrier density in the top ($n\sub{t}$) and bottom ($n\sub{b}$) layer, respectively. Black solid line corresponds to the total electron density ($n\sub{tot}$) in the bilayer \mos. At $\VBG = 5.4\unit{V}$ (red circle) same carrier density in top and bottom \mos layer is achieved.
}
\end{figure}

Figure~\ref{fig:fig2}(a) shows $\Delta R\sub{14,23}$ (color scale) as a function of $B$ applied perpendicular to the sample and \VBG, at $\VTG = 9\unit{V}$ and $T = 1.5\unit{K}$.
For $\VBG < 1.3\unit{V}$ (black arrow in \FigRefi{fig:fig2}{a}) the Landau level's evolution as a function of \VBG resembles the one of monolayer \mos~\cite{pisoni_interactions_2018}.
For $\VBG < -1.5\unit{V}$ only the Landau levels of the lower spin-orbit split $K_\uparrow$ and $K'_\downarrow$ bands are seen. As the electron density increases in this regime, we observe an alternating parity of the filling factor sequence [see filling factor sequences in \FigRefi{fig:fig2}{a}]. These results can be explained in an extended single-particle picture where the valley $g$ factor is density dependent, following the interpretation of previous works~\cite{lin_probing_2018,fallahazad_shubnikovchar21haas_2016, movva_density-dependent_2017, gustafsson_ambipolar_2017, larentis_large_2018}.

At $\VBG = -1.5\unit{V}$ there is a sudden change in the slope of the Landau fan related to the occupation of the higher spin-orbit split $K_\downarrow$ and $K'_\uparrow$ valleys.
Where the slope changes, the electron density is $n = 3.6 \times 10^{12} \psqcm$. Assuming two-fold valley-degeneracy and using the experimentally determined electron effective mass, we calculate the Fermi energy to be $E\sub{F} = 14 \mev$, in good agreement with the intrinsic spin-orbit interaction measured previously for $K$-valley electrons in monolayer \mos~\cite{pisoni_interactions_2018}. We would like to note that our results justify the assumptions in~\cite{lin_determining_2019} that bilayer \mos investigated in the right regime behaves as single-layer \mos with the caveat that the effective mass is different because of the dielectric environment.

The measured Landau level structure for $\VBG < 1.2\unit{V}$ fully agrees with our previous monolayer \mos results~\cite{pisoni_interactions_2018}.
For $\VBG$ crossing the voltage $1.2\unit{V}$ from below, we observe two important changes in the SdH oscillations compared to the monolayer system. First, the slope of the SdH oscillation minima vs \VBG that existed below this threshold changes sign from positive to negative. Second, an additional set of Landau levels appears [blue dashed lines in \FigRefi{fig:fig2}{a}].
At $\VBG = 3.4\unit{V}$ the slope of these secondary Landau levels changes by about a factor of 2, as indicated with white dashed lines in \FigRefi{fig:fig2}{a}.

To interpret these observations, we extract how the electron densities change as we tune \VBG. To determine the electron density of the individual layers and bands from the Landau fan diagram we generate a Fourier transform map of $\Delta R\sub{14,23}$ vs. $1/B$ for each \VBG value in \FigRefi{fig:fig2}{a} (see supplemental information).
The Fourier transform of the SdH oscillations shows multiple peaks in the amplitude spectrum as we increase \VBG. From these peaks we extract the electron density of the various spin-orbit split bands in bilayer \mos using $n = (g_\mathrm{v}e/h)\times f$, where $f$ is the frequency of the Fourier transform peaks and $g_\mathrm{v}=2$ accounts for the valley degeneracy.
The results of this procedure are shown in \FigRefi{fig:fig2}{b}.
For $\VBG < 1.2\unit{V}$ electrons populate only the top \mos layer where they occupy the lower (green dashed line) and upper (orange dashed line) spin-orbit split bands as we increase \VBG. 
At $\VBG = 1.2\unit{V}$ the bottom \mos layer starts to be populated (blue dashed line). The secondary Landau fan that appears at $\VBG = 1.2\unit{V}$ in \FigRefi{fig:fig2}{a} originates from the Landau levels of the electrons populating the bottom \mos layer. Beyond $\VBG = 1.2\unit{V}$ we see an increasing density in the bottom layer, whereas the top layer density starts to {\em decrease}. This density decrease is direct experimental evidence for the negative compressibility of the bottom layer at low densities~\cite{eisenstein_negative_1992,bello_density_1981,tanatar_ground_1989,kravchenko_evidence_1990}.
At $\VBG > 3.4\unit{V}$ the two valleys of the upper spin-orbit split bands in the bottom layer start to be populated (cyan dashed line).


%
\begin{figure}
\includegraphics[width=1\columnwidth]{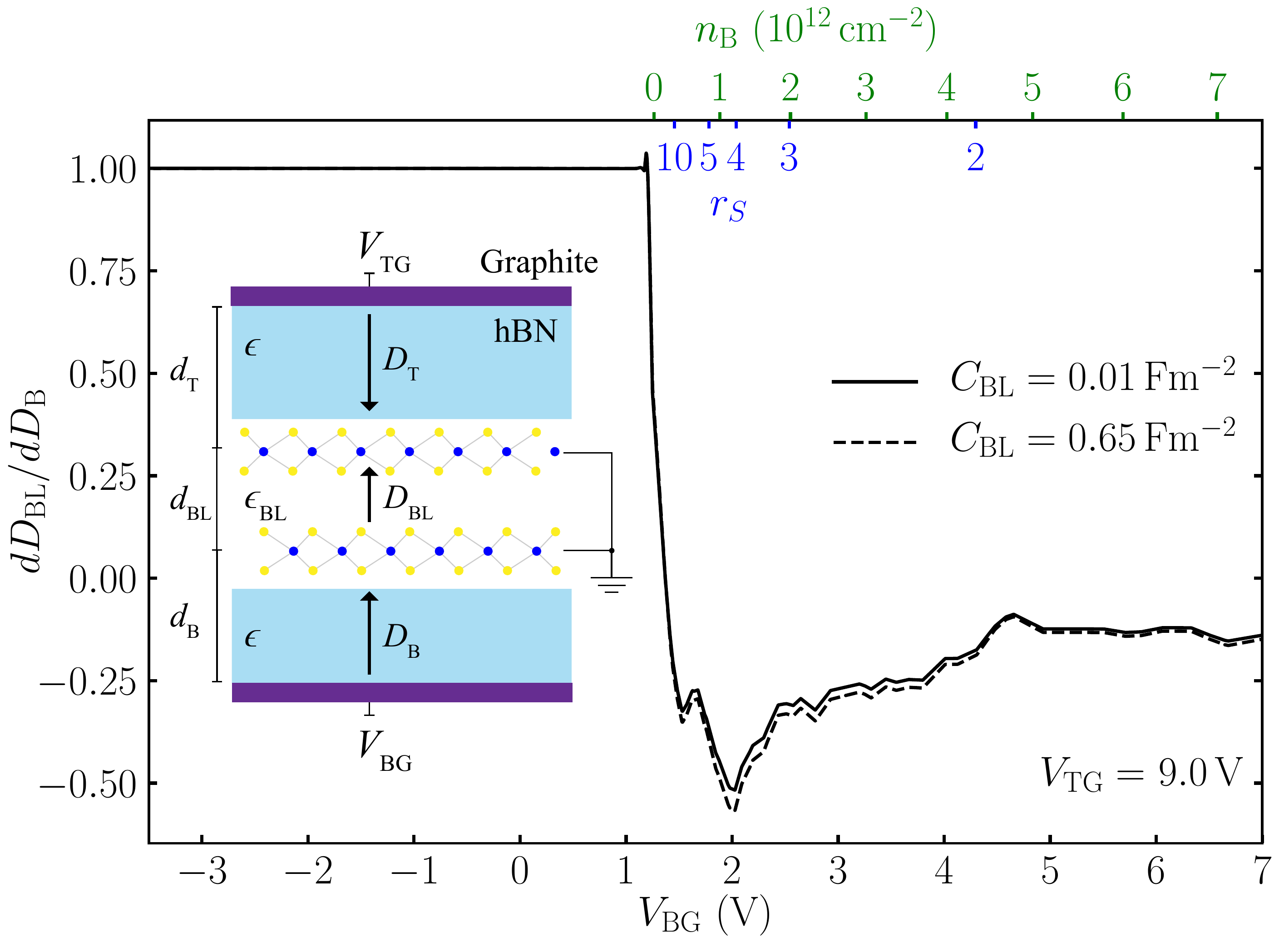}
\caption{\label{fig:fig3}
Sample A. Ratio of the electric displacement fields as a function of $\VBG$ at $T \approx 1.5\unit{K}$ and $\VTG=9\unit{V}$. The top green axis represents the electron density in the bottom layer ($n_\mathrm{B}$). The blue axis denotes the $r_S$ parameter that accounts for intralayer interactions in the bottom layer assuming in-plane dielectric constant $\approx 15.3$ ~\cite{laturia_dielectric_2018} and the measured effective mass of $0.6 m_\mathrm{e}$.
Inset: electrostatic model of our dual-gated bilayer \mos device.}
\end{figure}

In the following, we will quantify this negative compressibility effect in \mos based on the data in \FigRef{fig:fig2}.
To this end, we consider the electrostatic model schematically displayed in the inset of \FigRef{fig:fig3} consisting of three layers of different dielectric constants, in which electric displacement fields exist due to the applied voltages $\VTG$ and $\VBG$. The \mos bilayer is modeled as two grounded conducting planes of finite density of states with a geometric capacitance $\Cbl$ and a displacement field $\Dibl$ between them. It is our goal to express $d\Dibl/d\DiB$, i.e., the change in $\Dibl$ upon a change in the displacement field $\DiB$ between back gate and \mos, at constant top gate voltage in terms of the measured $\VBG$-dependent changes of the layer densities. This quantity allows us to directly compare the strength of the effect with the results obtained by Eisenstein {\em et al}~\cite{eisenstein_negative_1992} in the case of a GaAs double quantum well, and with the numerical results of Tanatar and Ceperley~\cite{tanatar_ground_1989}.

The model (see supplemental material for details) results in
\begin{equation*}
    \left.\frac{d\Dibl}{d\DiB}\right|_\VT = \frac{\Cbl}{\CB} \times 
\end{equation*}
\begin{equation}
    \times\frac{\CB\left(\CT + e \left.\frac{d\nT}{d\VB}\right|_\VT\right) - \CT e \left.\frac{d\nB}{d\VB}\right|_\VT}{\Cbl\left(\CT + e \left.\frac{d(\nT+\nB)}{d\VB}\right|_\VT \right)+\CT e \left.\frac{d\nB}{d\VB}\right|_\VT}\label{eq:ratio_D's},   
\end{equation}
where $\CT$ and $\CB$ are the geometric capacitances per unit area between \mos and top- and bottom-gate, respectively.
The quantities $\nT$ and $\nB$ are the measured total electron densities in the two layers shown in \FigRefi{fig:fig2}{b}. In the case of $\VBG<1.2\unit{V}$, where $\nB$ and its $\VBG$-derivative are zero, the displacement field ratio in eq.~\eqref{eq:ratio_D's} is exactly one. Negative compressibility in the region $\VB>1.2\unit{V}$ will manifest itself by $d\Dibl/d\DiB<0$.

Figure~\ref{fig:fig3} displays the result of applying eq.~\eqref{eq:ratio_D's} to the data in \FigRefi{fig:fig2}{b}. Two curves are shown, in which the quite uncertain value of $\Cbl$ takes on two plausible extreme values (see supplemental information for details). This shows that the result depends very little on the exact value of this parameter. A strong negative compressibility with $d\Dibl/\DiB\approx -0.5$ is seen at $\VBG=-2\unit{V}$ where $\nB=1\times 10^{12}\unit{cm^{-2}}$, roughly ten times stronger than the effect observed in Ref.~\cite{eisenstein_negative_1992}. To compare the value to the numerical results of Ref.~\cite{tanatar_ground_1989}, we have added an estimated scale bar of $r_\mathrm{s}$-values to the top axis in \FigRef{fig:fig3}. The negative compressibility values measured in our sample agree fairly well with the predictions of the numerical calculations at these $r_\mathrm{s}$-values.

Resolving individual layer electron densities in \FigRef{fig:fig2} indicates that the two \mos layers are weakly coupled. This observation is in contrast to Bernal stacked bilayer graphene, where the interlayer coupling of $\approx 0.4\unit{eV}$ strongly affects the energy momentum dispersion compared to monolayer graphene~\cite{zhang_determination_2008}.
Previous work probing hole transport in bilayer \wse reported an upper bound for the interlayer tunnel coupling of $\approx 19\unit{meV}$~\cite{fallahazad_shubnikovchar21haas_2016}.
In our results the interlayer coupling in the conduction band of bilayer \mos is not observable. We achieve same electron densities in both layers (red circle in \FigRefi{fig:fig2}{b}) for three different samples with no experimental evidence for interlayer coupling.
Band structure calculations~\cite{kormanyos_tunable_2018} reveal that strong interlayer hybridization in the conduction band of \mos occurs predominantly from orbitals which are responsible for the minima at the $Q$-point, which are not occupied in our samples. Conversely, weak interlayer hybridization is expected from the orbitals forming the $K$-valleys, which is consistent with our experimental observations.


At lower temperatures finer details of the Landau level structure are resolved.
In \FigRefi{fig:fig4}{a} we show $\Delta R\sub{14,23}$ as a function of $B$ and \VBG at $\VTG = 13.5\unit{V}$ and $T = 100\unit{mK}$ for sample B.
For $\VBG < 2\unit{V}$ (white dashed line) the bottom \mos layer is devoid of electrons and we only observe the Shubnikov-de Haas oscillations of the top layer. At 
$T = 100\unit{mK}$ we are able to resolve valley-spin polarized Landau levels originating from the lowest conduction band minima in the top layer.
The Landau level structure of the spin-valley coupled bands in the bottom layer appears for $\VBG \geq 2\unit{V}$.
Figures~\ref{fig:fig4}(b-c) show an enlargement of \FigRefi{fig:fig4}{a}. When only the top layer is populated [see \FigRefi{fig:fig4}{b}] we observe a pattern of avoided crossings, a signature of the coupling between the spin-valley polarized Landau levels of the lower and upper spin-orbit split bands~\cite{pisoni_interactions_2018}.
Figure~\ref{fig:fig4}(c) shows that Landau levels of electrons populating the two different layers cross each other, indicating weak coupling between the two electronic systems below our measurement resolution.

\begin{figure}[t!]
\includegraphics[width=1\columnwidth]{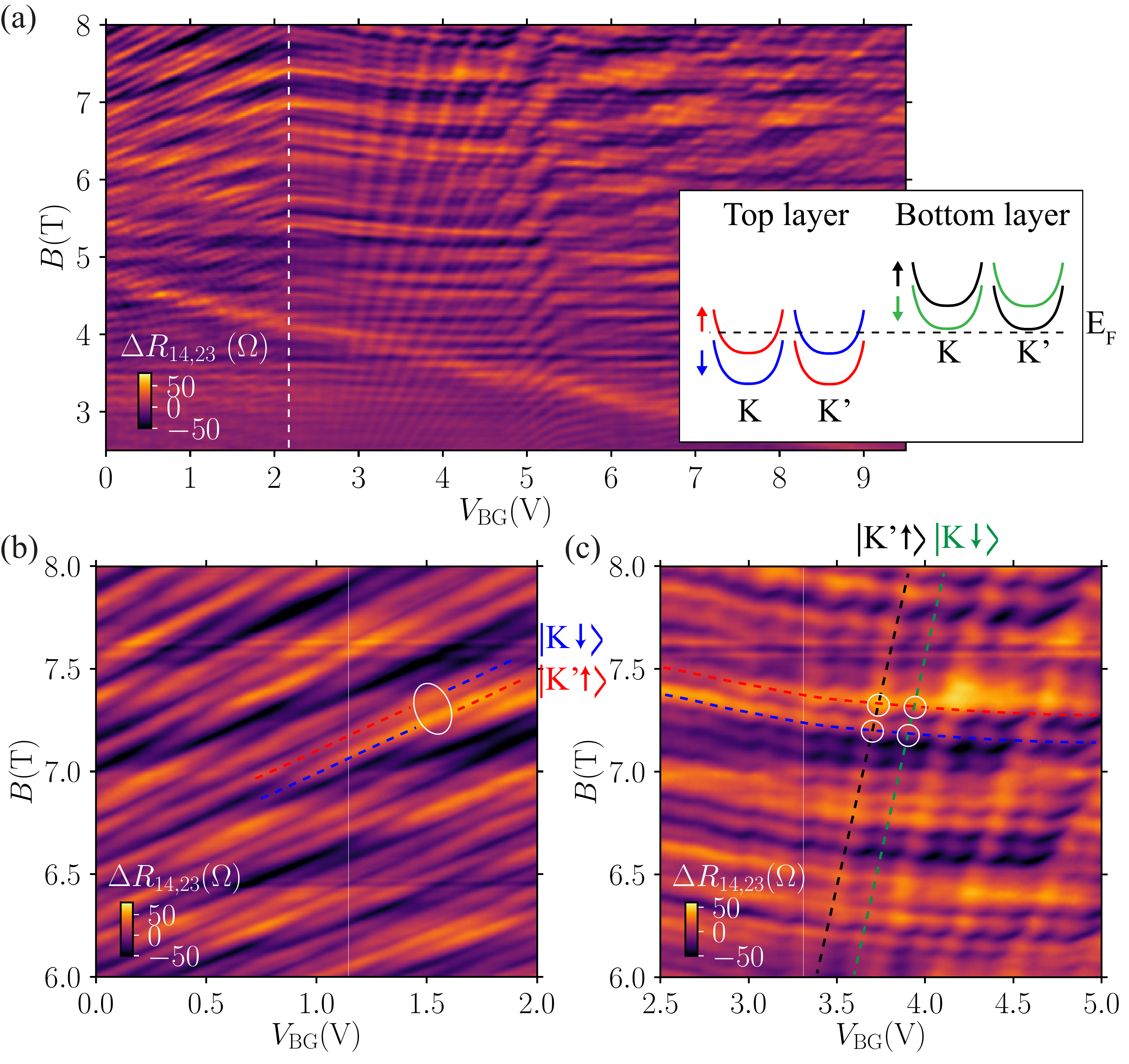}
\caption{\label{fig:fig4}
(a)~Sample B. Four-terminal resistance $\Delta R\sub{14,23}$ as a function of \VBG and magnetic field at $T\approx100\unit{mK}$ and $\VTG = 13.5\unit{V}$. At $\VBG = 2.2\unit{V}$ the bottom \mos layer is filled with electrons and SdH oscillations appear (white dashed lines). Inset: conduction band minima sketch at the $K$ and $K'$ point for top and bottom \mos layer. The horizontal black dashed line corresponds to the highest Fermi energy reached in the top layer before the bottom layer is occupied.
(b)~Avoided crossing patterns between spin-valley coupled LLs originating from the lower and upper spin-orbit split bands in the top \mos layer.
(c)~Crossings between LLs in the lower spin-orbit split bands originating from top and bottom \mos layers.
}
\end{figure}


In conclusion, we observe SdH oscillations at magnetic fields as low as $2\unit{T}$ at $T\approx100\unit{mK}$, testifying to the high-mobility of our dual-gated bilayer \mos devices.
We are able to measure spin-valley polarized LLs originating from the lower and upper spin-orbit split bands of $K$-valley electrons populating the top and bottom \mos layers. Our observations demonstrate that electrons in bilayer \mos behave like two independent electronic systems. The exchange interaction at the turn on of the two-dimensional electron gas in the bottom layer leads to the observation of a large negative compressibility.
Our work demonstrates fundamental electronic transport properties as well as the importance of interaction effects in pristine bilayer \mos. These results bear relevance for understanding electronic transport in twisted bilayer \tmds.

\begin{acknowledgments}
We thank Guido Burkard, Vladimir Falko, Andor Korm\'{a}nyos, Mansour Shayegan and Peter Rickhaus for fruitful discussions. We thank Peter M{\"a}rki as well as the FIRST staff for their technical support. We acknowledge financial support from ITN Spin-NANO Marie Sklodowska-Curie grant agreement no. 676108, the Graphene Flagship and the National Center of Competence in Research on Quantum Science and Technology (NCCR QSIT) funded by the Swiss National Science Foundation. 
Growth of hexagonal boron nitride crystals was supported by the Elemental Strategy Initiative conducted by the MEXT, Japan and JSPS KAKENHI Grant Numbers JP15K21722.
\end{acknowledgments}

\newpage
\bibliography{ref}

\begin{thebibliography}{33}%
\makeatletter
\providecommand \@ifxundefined [1]{%
 \@ifx{#1\undefined}
}%
\providecommand \@ifnum [1]{%
 \ifnum #1\expandafter \@firstoftwo
 \else \expandafter \@secondoftwo
 \fi
}%
\providecommand \@ifx [1]{%
 \ifx #1\expandafter \@firstoftwo
 \else \expandafter \@secondoftwo
 \fi
}%
\providecommand \natexlab [1]{#1}%
\providecommand \enquote  [1]{``#1''}%
\providecommand \bibnamefont  [1]{#1}%
\providecommand \bibfnamefont [1]{#1}%
\providecommand \citenamefont [1]{#1}%
\providecommand \href@noop [0]{\@secondoftwo}%
\providecommand \href [0]{\begingroup \@sanitize@url \@href}%
\providecommand \@href[1]{\@@startlink{#1}\@@href}%
\providecommand \@@href[1]{\endgroup#1\@@endlink}%
\providecommand \@sanitize@url [0]{\catcode `\\12\catcode `\$12\catcode
  `\&12\catcode `\#12\catcode `\^12\catcode `\_12\catcode `\%12\relax}%
\providecommand \@@startlink[1]{}%
\providecommand \@@endlink[0]{}%
\providecommand \url  [0]{\begingroup\@sanitize@url \@url }%
\providecommand \@url [1]{\endgroup\@href {#1}{\urlprefix }}%
\providecommand \urlprefix  [0]{URL }%
\providecommand \Eprint [0]{\href }%
\providecommand \doibase [0]{http://dx.doi.org/}%
\providecommand \selectlanguage [0]{\@gobble}%
\providecommand \bibinfo  [0]{\@secondoftwo}%
\providecommand \bibfield  [0]{\@secondoftwo}%
\providecommand \translation [1]{[#1]}%
\providecommand \BibitemOpen [0]{}%
\providecommand \bibitemStop [0]{}%
\providecommand \bibitemNoStop [0]{.\EOS\space}%
\providecommand \EOS [0]{\spacefactor3000\relax}%
\providecommand \BibitemShut  [1]{\csname bibitem#1\endcsname}%
\let\auto@bib@innerbib\@empty
\bibitem [{\citenamefont {Xiao}\ \emph {et~al.}(2012)\citenamefont {Xiao},
  \citenamefont {Liu}, \citenamefont {Feng}, \citenamefont {Xu},\ and\
  \citenamefont {Yao}}]{xiao_coupled_2012}%
  \BibitemOpen
  \bibfield  {author} {\bibinfo {author} {\bibfnamefont {D.}~\bibnamefont
  {Xiao}}, \bibinfo {author} {\bibfnamefont {G.-B.}\ \bibnamefont {Liu}},
  \bibinfo {author} {\bibfnamefont {W.}~\bibnamefont {Feng}}, \bibinfo {author}
  {\bibfnamefont {X.}~\bibnamefont {Xu}}, \ and\ \bibinfo {author}
  {\bibfnamefont {W.}~\bibnamefont {Yao}},\ }\href {\doibase
  10.1103/PhysRevLett.108.196802} {\bibfield  {journal} {\bibinfo  {journal}
  {Physical Review Letters}\ }\textbf {\bibinfo {volume} {108}},\ \bibinfo
  {pages} {196802} (\bibinfo {year} {2012})}\BibitemShut {NoStop}%
\bibitem [{\citenamefont {Mak}\ \emph {et~al.}(2012)\citenamefont {Mak},
  \citenamefont {He}, \citenamefont {Shan},\ and\ \citenamefont
  {Heinz}}]{mak_control_2012}%
  \BibitemOpen
  \bibfield  {author} {\bibinfo {author} {\bibfnamefont {K.~F.}\ \bibnamefont
  {Mak}}, \bibinfo {author} {\bibfnamefont {K.}~\bibnamefont {He}}, \bibinfo
  {author} {\bibfnamefont {J.}~\bibnamefont {Shan}}, \ and\ \bibinfo {author}
  {\bibfnamefont {T.~F.}\ \bibnamefont {Heinz}},\ }\href {\doibase
  10.1038/nnano.2012.96} {\bibfield  {journal} {\bibinfo  {journal} {Nature
  Nanotechnology}\ }\textbf {\bibinfo {volume} {7}},\ \bibinfo {pages} {494}
  (\bibinfo {year} {2012})}\BibitemShut {NoStop}%
\bibitem [{\citenamefont {Cao}\ \emph {et~al.}(2012)\citenamefont {Cao},
  \citenamefont {Wang}, \citenamefont {Han}, \citenamefont {Ye}, \citenamefont
  {Zhu}, \citenamefont {Shi}, \citenamefont {Niu}, \citenamefont {Tan},
  \citenamefont {Wang}, \citenamefont {Liu},\ and\ \citenamefont
  {Feng}}]{cao_valley-selective_2012}%
  \BibitemOpen
  \bibfield  {author} {\bibinfo {author} {\bibfnamefont {T.}~\bibnamefont
  {Cao}}, \bibinfo {author} {\bibfnamefont {G.}~\bibnamefont {Wang}}, \bibinfo
  {author} {\bibfnamefont {W.}~\bibnamefont {Han}}, \bibinfo {author}
  {\bibfnamefont {H.}~\bibnamefont {Ye}}, \bibinfo {author} {\bibfnamefont
  {C.}~\bibnamefont {Zhu}}, \bibinfo {author} {\bibfnamefont {J.}~\bibnamefont
  {Shi}}, \bibinfo {author} {\bibfnamefont {Q.}~\bibnamefont {Niu}}, \bibinfo
  {author} {\bibfnamefont {P.}~\bibnamefont {Tan}}, \bibinfo {author}
  {\bibfnamefont {E.}~\bibnamefont {Wang}}, \bibinfo {author} {\bibfnamefont
  {B.}~\bibnamefont {Liu}}, \ and\ \bibinfo {author} {\bibfnamefont
  {J.}~\bibnamefont {Feng}},\ }\href {\doibase 10.1038/ncomms1882} {\bibfield
  {journal} {\bibinfo  {journal} {Nature Communications}\ }\textbf {\bibinfo
  {volume} {3}},\ \bibinfo {pages} {887} (\bibinfo {year} {2012})}\BibitemShut
  {NoStop}%
\bibitem [{\citenamefont {Yao}\ \emph {et~al.}(2008)\citenamefont {Yao},
  \citenamefont {Xiao},\ and\ \citenamefont {Niu}}]{yao_valley-dependent_2008}%
  \BibitemOpen
  \bibfield  {author} {\bibinfo {author} {\bibfnamefont {W.}~\bibnamefont
  {Yao}}, \bibinfo {author} {\bibfnamefont {D.}~\bibnamefont {Xiao}}, \ and\
  \bibinfo {author} {\bibfnamefont {Q.}~\bibnamefont {Niu}},\ }\href {\doibase
  10.1103/PhysRevB.77.235406} {\bibfield  {journal} {\bibinfo  {journal}
  {Physical Review B}\ }\textbf {\bibinfo {volume} {77}},\ \bibinfo {pages}
  {235406} (\bibinfo {year} {2008})}\BibitemShut {NoStop}%
\bibitem [{\citenamefont {Wu}\ \emph {et~al.}(2013)\citenamefont {Wu},
  \citenamefont {Ross}, \citenamefont {Liu}, \citenamefont {Aivazian},
  \citenamefont {Jones}, \citenamefont {Fei}, \citenamefont {Zhu},
  \citenamefont {Xiao}, \citenamefont {Yao}, \citenamefont {Cobden},\ and\
  \citenamefont {Xu}}]{wu_electrical_2013}%
  \BibitemOpen
  \bibfield  {author} {\bibinfo {author} {\bibfnamefont {S.}~\bibnamefont
  {Wu}}, \bibinfo {author} {\bibfnamefont {J.~S.}\ \bibnamefont {Ross}},
  \bibinfo {author} {\bibfnamefont {G.-B.}\ \bibnamefont {Liu}}, \bibinfo
  {author} {\bibfnamefont {G.}~\bibnamefont {Aivazian}}, \bibinfo {author}
  {\bibfnamefont {A.}~\bibnamefont {Jones}}, \bibinfo {author} {\bibfnamefont
  {Z.}~\bibnamefont {Fei}}, \bibinfo {author} {\bibfnamefont {W.}~\bibnamefont
  {Zhu}}, \bibinfo {author} {\bibfnamefont {D.}~\bibnamefont {Xiao}}, \bibinfo
  {author} {\bibfnamefont {W.}~\bibnamefont {Yao}}, \bibinfo {author}
  {\bibfnamefont {D.}~\bibnamefont {Cobden}}, \ and\ \bibinfo {author}
  {\bibfnamefont {X.}~\bibnamefont {Xu}},\ }\href {\doibase 10.1038/nphys2524}
  {\bibfield  {journal} {\bibinfo  {journal} {Nature Physics}\ }\textbf
  {\bibinfo {volume} {9}},\ \bibinfo {pages} {149} (\bibinfo {year}
  {2013})}\BibitemShut {NoStop}%
\bibitem [{\citenamefont {Lee}\ \emph {et~al.}(2016)\citenamefont {Lee},
  \citenamefont {Mak},\ and\ \citenamefont {Shan}}]{lee_electrical_2016}%
  \BibitemOpen
  \bibfield  {author} {\bibinfo {author} {\bibfnamefont {J.}~\bibnamefont
  {Lee}}, \bibinfo {author} {\bibfnamefont {K.~F.}\ \bibnamefont {Mak}}, \ and\
  \bibinfo {author} {\bibfnamefont {J.}~\bibnamefont {Shan}},\ }\href {\doibase
  10.1038/nnano.2015.337} {\bibfield  {journal} {\bibinfo  {journal} {Nature
  Nanotechnology}\ }\textbf {\bibinfo {volume} {11}},\ \bibinfo {pages} {421}
  (\bibinfo {year} {2016})}\BibitemShut {NoStop}%
\bibitem [{\citenamefont {Lin}\ \emph {et~al.}(2018)\citenamefont {Lin},
  \citenamefont {Han}, \citenamefont {Piot}, \citenamefont {Wu}, \citenamefont
  {Xu}, \citenamefont {Long}, \citenamefont {An}, \citenamefont {Cheung},
  \citenamefont {Zheng}, \citenamefont {Plochocka}, \citenamefont {Maude},
  \citenamefont {Zhang},\ and\ \citenamefont {Wang}}]{lin_probing_2018}%
  \BibitemOpen
  \bibfield  {author} {\bibinfo {author} {\bibfnamefont {J.}~\bibnamefont
  {Lin}}, \bibinfo {author} {\bibfnamefont {T.}~\bibnamefont {Han}}, \bibinfo
  {author} {\bibfnamefont {B.~A.}\ \bibnamefont {Piot}}, \bibinfo {author}
  {\bibfnamefont {Z.}~\bibnamefont {Wu}}, \bibinfo {author} {\bibfnamefont
  {S.}~\bibnamefont {Xu}}, \bibinfo {author} {\bibfnamefont {G.}~\bibnamefont
  {Long}}, \bibinfo {author} {\bibfnamefont {L.}~\bibnamefont {An}}, \bibinfo
  {author} {\bibfnamefont {P.~K.~M.}\ \bibnamefont {Cheung}}, \bibinfo {author}
  {\bibfnamefont {P.-P.}\ \bibnamefont {Zheng}}, \bibinfo {author}
  {\bibfnamefont {P.}~\bibnamefont {Plochocka}}, \bibinfo {author}
  {\bibfnamefont {D.~K.}\ \bibnamefont {Maude}}, \bibinfo {author}
  {\bibfnamefont {F.}~\bibnamefont {Zhang}}, \ and\ \bibinfo {author}
  {\bibfnamefont {N.}~\bibnamefont {Wang}},\ }\href
  {http://arxiv.org/abs/1803.08007} {\bibfield  {journal} {\bibinfo  {journal}
  {arXiv:1803.08007 [cond-mat]}\ } (\bibinfo {year} {2018})},\ \bibinfo {note}
  {arXiv: 1803.08007}\BibitemShut {NoStop}%
\bibitem [{\citenamefont {Fallahazad}\ \emph {et~al.}(2016)\citenamefont
  {Fallahazad}, \citenamefont {Movva}, \citenamefont {Kim}, \citenamefont
  {Larentis}, \citenamefont {Taniguchi}, \citenamefont {Watanabe},
  \citenamefont {Banerjee},\ and\ \citenamefont
  {Tutuc}}]{fallahazad_shubnikovchar21haas_2016}%
  \BibitemOpen
  \bibfield  {author} {\bibinfo {author} {\bibfnamefont {B.}~\bibnamefont
  {Fallahazad}}, \bibinfo {author} {\bibfnamefont {H.~C.}\ \bibnamefont
  {Movva}}, \bibinfo {author} {\bibfnamefont {K.}~\bibnamefont {Kim}}, \bibinfo
  {author} {\bibfnamefont {S.}~\bibnamefont {Larentis}}, \bibinfo {author}
  {\bibfnamefont {T.}~\bibnamefont {Taniguchi}}, \bibinfo {author}
  {\bibfnamefont {K.}~\bibnamefont {Watanabe}}, \bibinfo {author}
  {\bibfnamefont {S.~K.}\ \bibnamefont {Banerjee}}, \ and\ \bibinfo {author}
  {\bibfnamefont {E.}~\bibnamefont {Tutuc}},\ }\href {\doibase
  10.1103/PhysRevLett.116.086601} {\bibfield  {journal} {\bibinfo  {journal}
  {Physical Review Letters}\ }\textbf {\bibinfo {volume} {116}},\ \bibinfo
  {pages} {086601} (\bibinfo {year} {2016})}\BibitemShut {NoStop}%
\bibitem [{\citenamefont {Movva}\ \emph {et~al.}(2017)\citenamefont {Movva},
  \citenamefont {Fallahazad}, \citenamefont {Kim}, \citenamefont {Larentis},
  \citenamefont {Taniguchi}, \citenamefont {Watanabe}, \citenamefont
  {Banerjee},\ and\ \citenamefont {Tutuc}}]{movva_density-dependent_2017}%
  \BibitemOpen
  \bibfield  {author} {\bibinfo {author} {\bibfnamefont {H.~C.}\ \bibnamefont
  {Movva}}, \bibinfo {author} {\bibfnamefont {B.}~\bibnamefont {Fallahazad}},
  \bibinfo {author} {\bibfnamefont {K.}~\bibnamefont {Kim}}, \bibinfo {author}
  {\bibfnamefont {S.}~\bibnamefont {Larentis}}, \bibinfo {author}
  {\bibfnamefont {T.}~\bibnamefont {Taniguchi}}, \bibinfo {author}
  {\bibfnamefont {K.}~\bibnamefont {Watanabe}}, \bibinfo {author}
  {\bibfnamefont {S.~K.}\ \bibnamefont {Banerjee}}, \ and\ \bibinfo {author}
  {\bibfnamefont {E.}~\bibnamefont {Tutuc}},\ }\href {\doibase
  10.1103/PhysRevLett.118.247701} {\bibfield  {journal} {\bibinfo  {journal}
  {Physical Review Letters}\ }\textbf {\bibinfo {volume} {118}},\ \bibinfo
  {pages} {247701} (\bibinfo {year} {2017})}\BibitemShut {NoStop}%
\bibitem [{\citenamefont {Larentis}\ \emph {et~al.}(2018)\citenamefont
  {Larentis}, \citenamefont {Movva}, \citenamefont {Fallahazad}, \citenamefont
  {Kim}, \citenamefont {Behroozi}, \citenamefont {Taniguchi}, \citenamefont
  {Watanabe}, \citenamefont {Banerjee},\ and\ \citenamefont
  {Tutuc}}]{larentis_large_2018}%
  \BibitemOpen
  \bibfield  {author} {\bibinfo {author} {\bibfnamefont {S.}~\bibnamefont
  {Larentis}}, \bibinfo {author} {\bibfnamefont {H.~C.~P.}\ \bibnamefont
  {Movva}}, \bibinfo {author} {\bibfnamefont {B.}~\bibnamefont {Fallahazad}},
  \bibinfo {author} {\bibfnamefont {K.}~\bibnamefont {Kim}}, \bibinfo {author}
  {\bibfnamefont {A.}~\bibnamefont {Behroozi}}, \bibinfo {author}
  {\bibfnamefont {T.}~\bibnamefont {Taniguchi}}, \bibinfo {author}
  {\bibfnamefont {K.}~\bibnamefont {Watanabe}}, \bibinfo {author}
  {\bibfnamefont {S.~K.}\ \bibnamefont {Banerjee}}, \ and\ \bibinfo {author}
  {\bibfnamefont {E.}~\bibnamefont {Tutuc}},\ }\href {\doibase
  10.1103/PhysRevB.97.201407} {\bibfield  {journal} {\bibinfo  {journal}
  {Physical Review B}\ }\textbf {\bibinfo {volume} {97}},\ \bibinfo {pages}
  {201407} (\bibinfo {year} {2018})}\BibitemShut {NoStop}%
\bibitem [{\citenamefont {Pisoni}\ \emph {et~al.}(2017)\citenamefont {Pisoni},
  \citenamefont {Lee}, \citenamefont {Overweg}, \citenamefont {Eich},
  \citenamefont {Simonet}, \citenamefont {Watanabe}, \citenamefont {Taniguchi},
  \citenamefont {Gorbachev}, \citenamefont {Ihn},\ and\ \citenamefont
  {Ensslin}}]{pisoni_gate-defined_2017}%
  \BibitemOpen
  \bibfield  {author} {\bibinfo {author} {\bibfnamefont {R.}~\bibnamefont
  {Pisoni}}, \bibinfo {author} {\bibfnamefont {Y.}~\bibnamefont {Lee}},
  \bibinfo {author} {\bibfnamefont {H.}~\bibnamefont {Overweg}}, \bibinfo
  {author} {\bibfnamefont {M.}~\bibnamefont {Eich}}, \bibinfo {author}
  {\bibfnamefont {P.}~\bibnamefont {Simonet}}, \bibinfo {author} {\bibfnamefont
  {K.}~\bibnamefont {Watanabe}}, \bibinfo {author} {\bibfnamefont
  {T.}~\bibnamefont {Taniguchi}}, \bibinfo {author} {\bibfnamefont
  {R.}~\bibnamefont {Gorbachev}}, \bibinfo {author} {\bibfnamefont
  {T.}~\bibnamefont {Ihn}}, \ and\ \bibinfo {author} {\bibfnamefont
  {K.}~\bibnamefont {Ensslin}},\ }\href {\doibase 10.1021/acs.nanolett.7b02186}
  {\bibfield  {journal} {\bibinfo  {journal} {Nano Letters}\ }\textbf {\bibinfo
  {volume} {17}},\ \bibinfo {pages} {5008} (\bibinfo {year}
  {2017})}\BibitemShut {NoStop}%
\bibitem [{\citenamefont {Pisoni}\ \emph
  {et~al.}(2018{\natexlab{a}})\citenamefont {Pisoni}, \citenamefont {Lei},
  \citenamefont {Back}, \citenamefont {Eich}, \citenamefont {Overweg},
  \citenamefont {Lee}, \citenamefont {Watanabe}, \citenamefont {Taniguchi},
  \citenamefont {Ihn},\ and\ \citenamefont
  {Ensslin}}]{pisoni_gate-tunable_2018}%
  \BibitemOpen
  \bibfield  {author} {\bibinfo {author} {\bibfnamefont {R.}~\bibnamefont
  {Pisoni}}, \bibinfo {author} {\bibfnamefont {Z.}~\bibnamefont {Lei}},
  \bibinfo {author} {\bibfnamefont {P.}~\bibnamefont {Back}}, \bibinfo {author}
  {\bibfnamefont {M.}~\bibnamefont {Eich}}, \bibinfo {author} {\bibfnamefont
  {H.}~\bibnamefont {Overweg}}, \bibinfo {author} {\bibfnamefont
  {Y.}~\bibnamefont {Lee}}, \bibinfo {author} {\bibfnamefont {K.}~\bibnamefont
  {Watanabe}}, \bibinfo {author} {\bibfnamefont {T.}~\bibnamefont {Taniguchi}},
  \bibinfo {author} {\bibfnamefont {T.}~\bibnamefont {Ihn}}, \ and\ \bibinfo
  {author} {\bibfnamefont {K.}~\bibnamefont {Ensslin}},\ }\href {\doibase
  10.1063/1.5021113} {\bibfield  {journal} {\bibinfo  {journal} {Applied
  Physics Letters}\ }\textbf {\bibinfo {volume} {112}},\ \bibinfo {pages}
  {123101} (\bibinfo {year} {2018}{\natexlab{a}})}\BibitemShut {NoStop}%
\bibitem [{\citenamefont {Pisoni}\ \emph
  {et~al.}(2018{\natexlab{b}})\citenamefont {Pisoni}, \citenamefont
  {Kormányos}, \citenamefont {Brooks}, \citenamefont {Lei}, \citenamefont
  {Back}, \citenamefont {Eich}, \citenamefont {Overweg}, \citenamefont {Lee},
  \citenamefont {Rickhaus}, \citenamefont {Watanabe}, \citenamefont
  {Taniguchi}, \citenamefont {Imamoglu}, \citenamefont {Burkard}, \citenamefont
  {Ihn},\ and\ \citenamefont {Ensslin}}]{pisoni_interactions_2018}%
  \BibitemOpen
  \bibfield  {author} {\bibinfo {author} {\bibfnamefont {R.}~\bibnamefont
  {Pisoni}}, \bibinfo {author} {\bibfnamefont {A.}~\bibnamefont {Kormányos}},
  \bibinfo {author} {\bibfnamefont {M.}~\bibnamefont {Brooks}}, \bibinfo
  {author} {\bibfnamefont {Z.}~\bibnamefont {Lei}}, \bibinfo {author}
  {\bibfnamefont {P.}~\bibnamefont {Back}}, \bibinfo {author} {\bibfnamefont
  {M.}~\bibnamefont {Eich}}, \bibinfo {author} {\bibfnamefont {H.}~\bibnamefont
  {Overweg}}, \bibinfo {author} {\bibfnamefont {Y.}~\bibnamefont {Lee}},
  \bibinfo {author} {\bibfnamefont {P.}~\bibnamefont {Rickhaus}}, \bibinfo
  {author} {\bibfnamefont {K.}~\bibnamefont {Watanabe}}, \bibinfo {author}
  {\bibfnamefont {T.}~\bibnamefont {Taniguchi}}, \bibinfo {author}
  {\bibfnamefont {A.}~\bibnamefont {Imamoglu}}, \bibinfo {author}
  {\bibfnamefont {G.}~\bibnamefont {Burkard}}, \bibinfo {author} {\bibfnamefont
  {T.}~\bibnamefont {Ihn}}, \ and\ \bibinfo {author} {\bibfnamefont
  {K.}~\bibnamefont {Ensslin}},\ }\href {\doibase
  10.1103/PhysRevLett.121.247701} {\bibfield  {journal} {\bibinfo  {journal}
  {Physical Review Letters}\ }\textbf {\bibinfo {volume} {121}},\ \bibinfo
  {pages} {247701} (\bibinfo {year} {2018}{\natexlab{b}})}\BibitemShut
  {NoStop}%
\bibitem [{\citenamefont {Ando}\ \emph {et~al.}(1982)\citenamefont {Ando},
  \citenamefont {Fowler},\ and\ \citenamefont {Stern}}]{ando_electronic_1982}%
  \BibitemOpen
  \bibfield  {author} {\bibinfo {author} {\bibfnamefont {T.}~\bibnamefont
  {Ando}}, \bibinfo {author} {\bibfnamefont {A.~B.}\ \bibnamefont {Fowler}}, \
  and\ \bibinfo {author} {\bibfnamefont {F.}~\bibnamefont {Stern}},\ }\href
  {\doibase 10.1103/RevModPhys.54.437} {\bibfield  {journal} {\bibinfo
  {journal} {Reviews of Modern Physics}\ }\textbf {\bibinfo {volume} {54}},\
  \bibinfo {pages} {437} (\bibinfo {year} {1982})}\BibitemShut {NoStop}%
\bibitem [{\citenamefont {Isihara}\ and\ \citenamefont
  {Smrcka}(1986)}]{isihara_density_1986}%
  \BibitemOpen
  \bibfield  {author} {\bibinfo {author} {\bibfnamefont {A.}~\bibnamefont
  {Isihara}}\ and\ \bibinfo {author} {\bibfnamefont {L.}~\bibnamefont
  {Smrcka}},\ }\href {\doibase 10.1088/0022-3719/19/34/015} {\bibfield
  {journal} {\bibinfo  {journal} {Journal of Physics C: Solid State Physics}\
  }\textbf {\bibinfo {volume} {19}},\ \bibinfo {pages} {6777} (\bibinfo {year}
  {1986})}\BibitemShut {NoStop}%
\bibitem [{\citenamefont {Pudalov}\ \emph {et~al.}(2014)\citenamefont
  {Pudalov}, \citenamefont {Gershenson},\ and\ \citenamefont
  {Kojima}}]{pudalov_probing_2014}%
  \BibitemOpen
  \bibfield  {author} {\bibinfo {author} {\bibfnamefont {V.~M.}\ \bibnamefont
  {Pudalov}}, \bibinfo {author} {\bibfnamefont {M.~E.}\ \bibnamefont
  {Gershenson}}, \ and\ \bibinfo {author} {\bibfnamefont {H.}~\bibnamefont
  {Kojima}},\ }\href {\doibase 10.1103/PhysRevB.90.075147} {\bibfield
  {journal} {\bibinfo  {journal} {Physical Review B}\ }\textbf {\bibinfo
  {volume} {90}},\ \bibinfo {pages} {075147} (\bibinfo {year}
  {2014})}\BibitemShut {NoStop}%
\bibitem [{\citenamefont {Zhang}\ and\ \citenamefont
  {Das~Sarma}(2005)}]{zhang_density-dependent_2005}%
  \BibitemOpen
  \bibfield  {author} {\bibinfo {author} {\bibfnamefont {Y.}~\bibnamefont
  {Zhang}}\ and\ \bibinfo {author} {\bibfnamefont {S.}~\bibnamefont
  {Das~Sarma}},\ }\href {\doibase 10.1103/PhysRevB.72.075308} {\bibfield
  {journal} {\bibinfo  {journal} {Physical Review B}\ }\textbf {\bibinfo
  {volume} {72}},\ \bibinfo {pages} {075308} (\bibinfo {year}
  {2005})}\BibitemShut {NoStop}%
\bibitem [{\citenamefont {Attaccalite}\ \emph {et~al.}(2002)\citenamefont
  {Attaccalite}, \citenamefont {Moroni}, \citenamefont {Gori-Giorgi},\ and\
  \citenamefont {Bachelet}}]{attaccalite_correlation_2002}%
  \BibitemOpen
  \bibfield  {author} {\bibinfo {author} {\bibfnamefont {C.}~\bibnamefont
  {Attaccalite}}, \bibinfo {author} {\bibfnamefont {S.}~\bibnamefont {Moroni}},
  \bibinfo {author} {\bibfnamefont {P.}~\bibnamefont {Gori-Giorgi}}, \ and\
  \bibinfo {author} {\bibfnamefont {G.~B.}\ \bibnamefont {Bachelet}},\ }\href
  {\doibase 10.1103/PhysRevLett.88.256601} {\bibfield  {journal} {\bibinfo
  {journal} {Physical Review Letters}\ }\textbf {\bibinfo {volume} {88}},\
  \bibinfo {pages} {256601} (\bibinfo {year} {2002})}\BibitemShut {NoStop}%
\bibitem [{\citenamefont {Cheiwchanchamnangij}\ and\ \citenamefont
  {Lambrecht}(2012)}]{cheiwchanchamnangij_quasiparticle_2012}%
  \BibitemOpen
  \bibfield  {author} {\bibinfo {author} {\bibfnamefont {T.}~\bibnamefont
  {Cheiwchanchamnangij}}\ and\ \bibinfo {author} {\bibfnamefont {W.~R.~L.}\
  \bibnamefont {Lambrecht}},\ }\href {\doibase 10.1103/PhysRevB.85.205302}
  {\bibfield  {journal} {\bibinfo  {journal} {Physical Review B}\ }\textbf
  {\bibinfo {volume} {85}},\ \bibinfo {pages} {205302} (\bibinfo {year}
  {2012})}\BibitemShut {NoStop}%
\bibitem [{\citenamefont {Ramasubramaniam}(2012)}]{ramasubramaniam_large_2012}%
  \BibitemOpen
  \bibfield  {author} {\bibinfo {author} {\bibfnamefont {A.}~\bibnamefont
  {Ramasubramaniam}},\ }\href {\doibase 10.1103/PhysRevB.86.115409} {\bibfield
  {journal} {\bibinfo  {journal} {Physical Review B}\ }\textbf {\bibinfo
  {volume} {86}},\ \bibinfo {pages} {115409} (\bibinfo {year}
  {2012})}\BibitemShut {NoStop}%
\bibitem [{\citenamefont {Molina-Sánchez}\ and\ \citenamefont
  {Wirtz}(2011)}]{molina-sanchez_phonons_2011}%
  \BibitemOpen
  \bibfield  {author} {\bibinfo {author} {\bibfnamefont {A.}~\bibnamefont
  {Molina-Sánchez}}\ and\ \bibinfo {author} {\bibfnamefont {L.}~\bibnamefont
  {Wirtz}},\ }\href {\doibase 10.1103/PhysRevB.84.155413} {\bibfield  {journal}
  {\bibinfo  {journal} {Physical Review B}\ }\textbf {\bibinfo {volume} {84}},\
  \bibinfo {pages} {155413} (\bibinfo {year} {2011})}\BibitemShut {NoStop}%
\bibitem [{\citenamefont {Laturia}\ \emph {et~al.}(2018)\citenamefont
  {Laturia}, \citenamefont {Put},\ and\ \citenamefont
  {Vandenberghe}}]{laturia_dielectric_2018}%
  \BibitemOpen
  \bibfield  {author} {\bibinfo {author} {\bibfnamefont {A.}~\bibnamefont
  {Laturia}}, \bibinfo {author} {\bibfnamefont {M.~L. V.~d.}\ \bibnamefont
  {Put}}, \ and\ \bibinfo {author} {\bibfnamefont {W.~G.}\ \bibnamefont
  {Vandenberghe}},\ }\href {\doibase 10.1038/s41699-018-0050-x} {\bibfield
  {journal} {\bibinfo  {journal} {npj 2D Materials and Applications}\ }\textbf
  {\bibinfo {volume} {2}},\ \bibinfo {pages} {6} (\bibinfo {year}
  {2018})}\BibitemShut {NoStop}%
\bibitem [{\citenamefont {Okamoto}\ \emph {et~al.}(1999)\citenamefont
  {Okamoto}, \citenamefont {Hosoya}, \citenamefont {Kawaji},\ and\
  \citenamefont {Yagi}}]{okamoto_spin_1999}%
  \BibitemOpen
  \bibfield  {author} {\bibinfo {author} {\bibfnamefont {T.}~\bibnamefont
  {Okamoto}}, \bibinfo {author} {\bibfnamefont {K.}~\bibnamefont {Hosoya}},
  \bibinfo {author} {\bibfnamefont {S.}~\bibnamefont {Kawaji}}, \ and\ \bibinfo
  {author} {\bibfnamefont {A.}~\bibnamefont {Yagi}},\ }\href {\doibase
  10.1103/PhysRevLett.82.3875} {\bibfield  {journal} {\bibinfo  {journal}
  {Physical Review Letters}\ }\textbf {\bibinfo {volume} {82}},\ \bibinfo
  {pages} {3875} (\bibinfo {year} {1999})}\BibitemShut {NoStop}%
\bibitem [{\citenamefont {Shashkin}\ \emph {et~al.}(2001)\citenamefont
  {Shashkin}, \citenamefont {Kravchenko}, \citenamefont {Dolgopolov},\ and\
  \citenamefont {Klapwijk}}]{shashkin_indication_2001}%
  \BibitemOpen
  \bibfield  {author} {\bibinfo {author} {\bibfnamefont {A.~A.}\ \bibnamefont
  {Shashkin}}, \bibinfo {author} {\bibfnamefont {S.~V.}\ \bibnamefont
  {Kravchenko}}, \bibinfo {author} {\bibfnamefont {V.~T.}\ \bibnamefont
  {Dolgopolov}}, \ and\ \bibinfo {author} {\bibfnamefont {T.~M.}\ \bibnamefont
  {Klapwijk}},\ }\href {\doibase 10.1103/PhysRevLett.87.086801} {\bibfield
  {journal} {\bibinfo  {journal} {Physical Review Letters}\ }\textbf {\bibinfo
  {volume} {87}},\ \bibinfo {pages} {086801} (\bibinfo {year}
  {2001})}\BibitemShut {NoStop}%
\bibitem [{\citenamefont {Vakili}\ \emph {et~al.}(2004)\citenamefont {Vakili},
  \citenamefont {Shkolnikov}, \citenamefont {Tutuc}, \citenamefont
  {De~Poortere},\ and\ \citenamefont {Shayegan}}]{vakili_spin_2004}%
  \BibitemOpen
  \bibfield  {author} {\bibinfo {author} {\bibfnamefont {K.}~\bibnamefont
  {Vakili}}, \bibinfo {author} {\bibfnamefont {Y.~P.}\ \bibnamefont
  {Shkolnikov}}, \bibinfo {author} {\bibfnamefont {E.}~\bibnamefont {Tutuc}},
  \bibinfo {author} {\bibfnamefont {E.~P.}\ \bibnamefont {De~Poortere}}, \ and\
  \bibinfo {author} {\bibfnamefont {M.}~\bibnamefont {Shayegan}},\ }\href
  {\doibase 10.1103/PhysRevLett.92.226401} {\bibfield  {journal} {\bibinfo
  {journal} {Physical Review Letters}\ }\textbf {\bibinfo {volume} {92}},\
  \bibinfo {pages} {226401} (\bibinfo {year} {2004})}\BibitemShut {NoStop}%
\bibitem [{\citenamefont {Gustafsson}\ \emph {et~al.}(2017)\citenamefont
  {Gustafsson}, \citenamefont {Yankowitz}, \citenamefont {Forsythe},
  \citenamefont {Rhodes}, \citenamefont {Watanabe}, \citenamefont {Taniguchi},
  \citenamefont {Hone}, \citenamefont {Zhu},\ and\ \citenamefont
  {Dean}}]{gustafsson_ambipolar_2017}%
  \BibitemOpen
  \bibfield  {author} {\bibinfo {author} {\bibfnamefont {M.~V.}\ \bibnamefont
  {Gustafsson}}, \bibinfo {author} {\bibfnamefont {M.}~\bibnamefont
  {Yankowitz}}, \bibinfo {author} {\bibfnamefont {C.}~\bibnamefont {Forsythe}},
  \bibinfo {author} {\bibfnamefont {D.}~\bibnamefont {Rhodes}}, \bibinfo
  {author} {\bibfnamefont {K.}~\bibnamefont {Watanabe}}, \bibinfo {author}
  {\bibfnamefont {T.}~\bibnamefont {Taniguchi}}, \bibinfo {author}
  {\bibfnamefont {J.}~\bibnamefont {Hone}}, \bibinfo {author} {\bibfnamefont
  {X.}~\bibnamefont {Zhu}}, \ and\ \bibinfo {author} {\bibfnamefont {C.~R.}\
  \bibnamefont {Dean}},\ }\href {http://arxiv.org/abs/1707.08083} {\bibfield
  {journal} {\bibinfo  {journal} {arXiv:1707.08083 [cond-mat]}\ } (\bibinfo
  {year} {2017})},\ \bibinfo {note} {arXiv: 1707.08083}\BibitemShut {NoStop}%
\bibitem [{\citenamefont {Lin}\ \emph {et~al.}(2019)\citenamefont {Lin},
  \citenamefont {Han}, \citenamefont {Piot}, \citenamefont {Wu}, \citenamefont
  {Xu}, \citenamefont {Long}, \citenamefont {An}, \citenamefont {Cheung},
  \citenamefont {Zheng}, \citenamefont {Plochocka}, \citenamefont {Dai},
  \citenamefont {Maude}, \citenamefont {Zhang},\ and\ \citenamefont
  {Wang}}]{lin_determining_2019}%
  \BibitemOpen
  \bibfield  {author} {\bibinfo {author} {\bibfnamefont {J.}~\bibnamefont
  {Lin}}, \bibinfo {author} {\bibfnamefont {T.}~\bibnamefont {Han}}, \bibinfo
  {author} {\bibfnamefont {B.~A.}\ \bibnamefont {Piot}}, \bibinfo {author}
  {\bibfnamefont {Z.}~\bibnamefont {Wu}}, \bibinfo {author} {\bibfnamefont
  {S.}~\bibnamefont {Xu}}, \bibinfo {author} {\bibfnamefont {G.}~\bibnamefont
  {Long}}, \bibinfo {author} {\bibfnamefont {L.}~\bibnamefont {An}}, \bibinfo
  {author} {\bibfnamefont {P.}~\bibnamefont {Cheung}}, \bibinfo {author}
  {\bibfnamefont {P.-P.}\ \bibnamefont {Zheng}}, \bibinfo {author}
  {\bibfnamefont {P.}~\bibnamefont {Plochocka}}, \bibinfo {author}
  {\bibfnamefont {X.}~\bibnamefont {Dai}}, \bibinfo {author} {\bibfnamefont
  {D.~K.}\ \bibnamefont {Maude}}, \bibinfo {author} {\bibfnamefont
  {F.}~\bibnamefont {Zhang}}, \ and\ \bibinfo {author} {\bibfnamefont
  {N.}~\bibnamefont {Wang}},\ }\href {\doibase 10.1021/acs.nanolett.8b04731}
  {\bibfield  {journal} {\bibinfo  {journal} {Nano Letters}\ }\textbf {\bibinfo
  {volume} {19}},\ \bibinfo {pages} {1736} (\bibinfo {year}
  {2019})}\BibitemShut {NoStop}%
\bibitem [{\citenamefont {Eisenstein}\ \emph {et~al.}(1992)\citenamefont
  {Eisenstein}, \citenamefont {Pfeiffer},\ and\ \citenamefont
  {West}}]{eisenstein_negative_1992}%
  \BibitemOpen
  \bibfield  {author} {\bibinfo {author} {\bibfnamefont {J.~P.}\ \bibnamefont
  {Eisenstein}}, \bibinfo {author} {\bibfnamefont {L.~N.}\ \bibnamefont
  {Pfeiffer}}, \ and\ \bibinfo {author} {\bibfnamefont {K.~W.}\ \bibnamefont
  {West}},\ }\href {\doibase 10.1103/PhysRevLett.68.674} {\bibfield  {journal}
  {\bibinfo  {journal} {Physical Review Letters}\ }\textbf {\bibinfo {volume}
  {68}},\ \bibinfo {pages} {674} (\bibinfo {year} {1992})}\BibitemShut
  {NoStop}%
\bibitem [{\citenamefont {Bello}\ \emph {et~al.}(1981)\citenamefont {Bello},
  \citenamefont {Levin},\ and\ \citenamefont {Shklovskr}}]{bello_density_1981}%
  \BibitemOpen
  \bibfield  {author} {\bibinfo {author} {\bibfnamefont {M.~S.}\ \bibnamefont
  {Bello}}, \bibinfo {author} {\bibfnamefont {E.~I.}\ \bibnamefont {Levin}}, \
  and\ \bibinfo {author} {\bibfnamefont {B.~I.}\ \bibnamefont {Shklovskr}},\
  }\href@noop {} {\ ,\ \bibinfo {pages} {8} (\bibinfo {year}
  {1981})}\BibitemShut {NoStop}%
\bibitem [{\citenamefont {Tanatar}\ and\ \citenamefont
  {Ceperley}(1989)}]{tanatar_ground_1989}%
  \BibitemOpen
  \bibfield  {author} {\bibinfo {author} {\bibfnamefont {B.}~\bibnamefont
  {Tanatar}}\ and\ \bibinfo {author} {\bibfnamefont {D.~M.}\ \bibnamefont
  {Ceperley}},\ }\href {\doibase 10.1103/PhysRevB.39.5005} {\bibfield
  {journal} {\bibinfo  {journal} {Physical Review B}\ }\textbf {\bibinfo
  {volume} {39}},\ \bibinfo {pages} {5005} (\bibinfo {year}
  {1989})}\BibitemShut {NoStop}%
\bibitem [{\citenamefont {Kravchenko}\ \emph {et~al.}(1990)\citenamefont
  {Kravchenko}, \citenamefont {Rinberg}, \citenamefont {Semenchinsky},\ and\
  \citenamefont {Pudalov}}]{kravchenko_evidence_1990}%
  \BibitemOpen
  \bibfield  {author} {\bibinfo {author} {\bibfnamefont {S.~V.}\ \bibnamefont
  {Kravchenko}}, \bibinfo {author} {\bibfnamefont {D.~A.}\ \bibnamefont
  {Rinberg}}, \bibinfo {author} {\bibfnamefont {S.~G.}\ \bibnamefont
  {Semenchinsky}}, \ and\ \bibinfo {author} {\bibfnamefont {V.~M.}\
  \bibnamefont {Pudalov}},\ }\href {\doibase 10.1103/PhysRevB.42.3741}
  {\bibfield  {journal} {\bibinfo  {journal} {Physical Review B}\ }\textbf
  {\bibinfo {volume} {42}},\ \bibinfo {pages} {3741} (\bibinfo {year}
  {1990})}\BibitemShut {NoStop}%
\bibitem [{\citenamefont {Zhang}\ \emph {et~al.}(2008)\citenamefont {Zhang},
  \citenamefont {Li}, \citenamefont {Basov}, \citenamefont {Fogler},
  \citenamefont {Hao},\ and\ \citenamefont
  {Martin}}]{zhang_determination_2008}%
  \BibitemOpen
  \bibfield  {author} {\bibinfo {author} {\bibfnamefont {L.~M.}\ \bibnamefont
  {Zhang}}, \bibinfo {author} {\bibfnamefont {Z.~Q.}\ \bibnamefont {Li}},
  \bibinfo {author} {\bibfnamefont {D.~N.}\ \bibnamefont {Basov}}, \bibinfo
  {author} {\bibfnamefont {M.~M.}\ \bibnamefont {Fogler}}, \bibinfo {author}
  {\bibfnamefont {Z.}~\bibnamefont {Hao}}, \ and\ \bibinfo {author}
  {\bibfnamefont {M.~C.}\ \bibnamefont {Martin}},\ }\href {\doibase
  10.1103/PhysRevB.78.235408} {\bibfield  {journal} {\bibinfo  {journal}
  {Physical Review B}\ }\textbf {\bibinfo {volume} {78}},\ \bibinfo {pages}
  {235408} (\bibinfo {year} {2008})}\BibitemShut {NoStop}%
\bibitem [{\citenamefont {Kormányos}\ \emph {et~al.}()\citenamefont
  {Kormányos}, \citenamefont {Zólyomi}, \citenamefont {Fal'ko},\ and\
  \citenamefont {Burkard}}]{kormanyos_tunable_2018}%
  \BibitemOpen
  \bibfield  {author} {\bibinfo {author} {\bibfnamefont {A.}~\bibnamefont
  {Kormányos}}, \bibinfo {author} {\bibfnamefont {V.}~\bibnamefont
  {Zólyomi}}, \bibinfo {author} {\bibfnamefont {V.~I.}\ \bibnamefont
  {Fal'ko}}, \ and\ \bibinfo {author} {\bibfnamefont {G.}~\bibnamefont
  {Burkard}},\ }\href {\doibase 10.1103/PhysRevB.98.035408} {\bibfield
  {journal} {\bibinfo  {journal} {Physical Review B}\ }\textbf {\bibinfo
  {volume} {98}},\ \bibinfo {pages} {035408 (2018) + private
  communication}}\BibitemShut {NoStop}%
\end{thebibliography}%

\end{document}